\documentclass[10pt]{scrartcl}
\usepackage{chemformula} 
\usepackage[T1]{fontenc} 
\usepackage[numbers,sort&compress]{natbib}
\usepackage{graphicx}
\usepackage[english]{babel}
\usepackage{url}
\usepackage{lipsum}
\usepackage{color}
\usepackage{amsmath}
\usepackage{appendix}
\usepackage{amssymb}

\title{ \LARGE \centering Combining SchNet and SHARC: The SchNarc machine learning approach for excited-state dynamics\\~\\ \Large Julia Westermayr,$^1$ Michael Gastegger,$^2$ and Philipp Marquetand$^{1,3}$\\~\\ \small $^1$University of Vienna, Institute of Theoretical Chemistry, W\"{a}hringer Str. 17, 1090 Vienna, Austria\\ $^2$Machine Learning Group, Technical University of Berlin, 10587 Berlin, Germany\\
$^3$Vienna Research Platform on Accelerating Photoreaction Discovery, University of Vienna, W\"ahringer Str. 17, 1090 Vienna, Austria.\\
E-mail: philipp.marquetand@univie.ac.at; michael.gastegger@tu-berlin.de
}
\date{}

\begin{document}
\maketitle

\begin{abstract}
In recent years, deep learning has become a part of our everyday life and is revolutionizing quantum chemistry as well. In this work, we show how deep learning can be used to advance the research field of photochemistry by learning all important properties for photodynamics simulations. The properties are multiple energies, forces, nonadiabatic couplings and spin-orbit couplings. The nonadiabatic couplings are learned in a phase-free manner as derivatives of a virtually constructed property by the deep learning model, which guarantees rotational covariance. Additionally, an approximation for nonadiabatic couplings is introduced, based on the potentials, their gradients and Hessians. As deep-learning method, we employ SchNet extended for multiple electronic states. In combination with the molecular dynamics program SHARC, our approach termed SchNarc is tested on a model system and two realistic polyatomic molecules and paves the way towards efficient photodynamics simulations of complex systems.
\end{abstract}
\newpage

Excited-state dynamics simulations are powerful tools to predict, understand and explain photo-induced processes, especially in combination with experimental studies. Examples of photo-induced processes range from photosynthesis, DNA photodamage as the starting point of skin cancer, to processes that enable our vision~\cite{Cerullo2002S,Schultz2004S,Schreier2007S,Rauer2016JACS,Romero2017N}. As they are part of our everyday lives, their understanding can help to unravel fundamental processes of nature and to advance several research fields, such as photovoltaics~\cite{mathew2014NC,Bartok2017SA}, photocatalysis~\cite{Sanchez-Lengeling2018S} or photosensitive drug design~\cite{Ahmad2016IJP}.

Since the full quantum mechanical treatment of molecules remains challenging, exact quantum dynamics simulations are limited to systems containing only a couple of atoms, even if fitted potential energy surfaces (PESs) are used~\cite{Koeppel84ACP,Koeppel2001JCP,Worth2004ARPC,Bowman2008MP,Meyer2009,Alborzpour2016JCP,Richings2017CPL,Liu2017SR,Richings2017CPL,Richings2018JCP,Williams2018JCP,Xie2018JCP,Guan2019PCCP,Richings2019JCTC,Polyak2019JCP,Guan2019JCP,Wang2019JPCA,Guan2020JCTC}. In order to treat larger systems in full dimensions, i.e., systems with up to 100s of atoms, and on long time scales, i.e., in the range of several 100 picoseconds, excited-state machine learning (ML) molecular dynamics (MD), where the ML model is trained on quantum chemistry data, has evolved as a promising tool in the last couple of years~\cite{Behler2008PRB,Carbogno2010PRB,Hu2018JPCL,Dral2018JPCL,Chen2018JPCL,Westermayr2019CS,Westermayr2019arXiv}. 

Such nonadiabatic MLMD simulations are in many senses analog to excited-state ab initio molecular dynamics simulations. The only difference is that the costly electronic structure calculations are mostly replaced by a ML model, providing quantum properties like the PESs and the corresponding forces. The nuclei are assumed to move classically on those PESs. This mixed quantum-classical dynamics approach allows for a very fast on-the-fly evaluation of the necessary properties at the geometries visited during the dynamics simulations. 

In order to account for nonadiabatic effects, i.e., transitions from one state to another, further approximations have to be introduced~\cite{Ibele2019MP}. One method, which is frequently used to account for such transitions, is the surface-hopping method originally developed by Tully~\cite{Tully1990JCP}. A popular extension for this method including not only nonadiabatic couplings (NACs) but also other couplings, e.g., spin-orbit couplings (SOCs), is the SHARC (surface hopping including arbitrary couplings) approach~\cite{Richter2011JCTC,Mai2018WCMS,sharc-md2}. 
Importantly, NACs, also called derivative couplings, are used to determine the hopping directions and probabilities between states of same spin multiplicity~\cite{Doltsinis2006NIC,Richter2011JCTC,Mai2018WCMS,Ha2018JPCL,Mai2020ACIE}. The NAC vector (denoted as $C^{\text{NAC}}$) between two states, i and j, can be computed as~\cite{Baer2002PR,Lischka2004JCP,Doltsinis2006NIC} 
\begin{equation}
    \label{eq:nac5}
        C^{\text{NAC}}_{ij} \approx \langle  \Psi_i \mid \frac{\partial}{\partial \mathbf{R}}\Psi_j \rangle = \frac{1}{E_i-E_j}{\langle \Psi_i \mid \frac{\partial H_{el}}{\partial \mathbf{R}}\mid \Psi_j \rangle }   ~~~\text{for}~  i \neq j,
\end{equation}
where the second-order derivatives are neglected. As a further difficulty, NACs are often missing from quantum chemistry implementations and are thus often approximated~\cite{Rubbmark1981PRA,Hagedorn1991CMP,Wittig2005JPCB,Zhu1995PRL,Zhu2001JCP,Zhu2002JCP,Kondorskiy2004JCP,Oloyede2006JCP,Ishida2017IRPC}. SOCs (denoted as $C^{\text{SOC}}$) are present between states of different spin multiplicity, 
\begin{equation}
C^{\text{SOC}}_{ij}=\langle \Psi_i \mid \hat{H}^{SO}\mid \Psi_j \rangle 
\end{equation}
and determine the rate of intersystem crossing. They are obtained as off-diagonal elements of the Hamiltonian matrix in standard electronic-structure calculations~\cite{Mai2018WCMS,Granucci2012JCP}.

Most of the recent studies involving ML dynamics deal with  ground-state MD simulations, see e.g. Refs.~\cite{Bartok2010PRL,Li2015PRL,Gastegger2015JCTC,Rupp2015IJQC,Behler2016JCP,Gastegger2016JCP,Gastegger2017CS,Deringer2017PRB,Botu2017JPCC,Smith2017CS,Behler2017ACIE,Zong2018npjCM,Bartok2018PRX,Chmiela2018NC,Imbalzano2018JCP,Zhang2018NIPS,Zhang2018PRL,Chan2019JPCC,Christensen2019JCP,Wang2019JCTC,Chmiela2019CPC,Carleo2019arXiv,Krems2019PCCP,Deringer2019AM,Schuett2019NC}, where one of the most promising ML models 
is SchNet~\cite{Schuett2018JCP,Schuett2019JCTC}, a deep continuous-filter convolutional-layer neural network. 

Only a small but quickly increasing number of studies deal with the treatment of excited states and their properties using ML~\cite{Behler2008PRB,Carbogno2010PRB,Alborzpour2016JCP,Haese2016CS,Richings2017CPL,Liu2017SR,Richings2017CPL,Richings2018JCP,Chen2018JPCL,Williams2018JCP,Xie2018JCP,Guan2019PCCP,Richings2019JCTC,Polyak2019JCP,Westermayr2019arXiv,Guan2019JCP,Wang2019JPCA,Guan2020JCTC}. An arising difficulty compared to ground state energies and properties is that not only one, but several PESs as well as the couplings between them have to be taken into account. Additionally, the learning of couplings proves challenging due to the fact that properties resulting from electronic wave functions of two different states, $\Psi_i$ and $\Psi_j$, have their sign dependent on the phase of the wave functions~\cite{Akimov2018JPCL,Bellonzi2019JCP,Westermayr2019CS,Westermayr2019arXiv}.
Since the wave function phase is not uniquely defined in quantum chemistry calculations, random phase jumps occur, leading to sign jumps of the coupling values along a reaction path. Hence, the couplings can not be learned directly as obtained from a quantum chemistry calculation. An option is to use a phase correction algorithm to pre-process data and remove these random phase jumps. Assuming that the effect of the Berry phase remains minor on the training set, smooth properties are obtained that are learnable by ML models~\cite{Westermayr2019CS}. However, this approach is expensive and many quantum chemistry reference computations are necessary to generate the training set. In cases of large poly-atomic molecules with many close lying energetic states, this approach might even be infeasible. 

The aim of this letter is to provide a framework to carry out efficient excited-state MLMD simulations and to combine two popular methods for this purpose: the SHARC approach for photodynamics with states of different multiplicity and SchNet to efficiently and accurately fit potential energies and other molecular properties. We call this combination the SchNarc approach and adapted SchNet for the treatment of excited state potentials, their forces and couplings for this purpose. The SchNarc approach can overcome the current limitations of existing MLMD simulations for excited states by allowing (i) a phase-free training to omit the costly pre-processing of raw quantum chemistry data and, to treat (ii) rotationally covariant NACs, which can either be trained, or (iii) alternatively be approximated from only ML potentials, their gradients, and Hessians, and to treat (iv) SOCs.

To validate the phase-free training, a new loss function termed \textit{phase-less loss function} is developed and tested for the methylenimmonium cation, CH$_2$NH$_2^+$, of which we take a phase corrected training set from Ref.~\cite{Westermayr2019CS}. Using the same level of theory (MR-CISD(6,4)/aug-cc-pVDZ) with the program COLUMBUS~\cite{Lischka2001PCCP}, the training set is recomputed without applying phase correction to train ML models also on raw data obtained directly from quantum chemistry programs. The ML models are trained on energies, gradients, and NACs for three singlet states using 3,000 data points. This molecular system features singlet-only dynamics and ultrafast transitions.

The phase-less loss is based on the standard L$_2$ loss, but here, the squared error of the predicted properties is computed $2^{N_S-1}$-times with N$_S$ being the total number states. The value of each property, $L_P$ (i.e. $L_{SOC}$ and $L_{NAC}$), that enters the loss function is the minimum function of all possible squared errors $\varepsilon^k_{P}$:

\begin{equation}
    \label{eq:lph}
    L_P = \text{min}\left(\{\varepsilon^{k}_{P}\}\right) ~~ \text{with} ~ 0 \leq k \leq 2^{N_S-1}
    \end{equation}
    with 
    \begin{equation}
      \varepsilon^{k}_{P} =  \left\{
\begin{array}{ll}
\frac{1}{N_S^2}\sum_i^{N_S}\sum_{j\neq i}^{N_S}\frac{1}{N_A} \sum_m^{N_A} \mid\mid P^{QC}_{ij,m} - P^{ML}_{ij,m} \cdot p_i^{k} \cdot p_j^{k} \mid\mid^2  & \text{if}~dim(\mathbf{P})\geq 3\\
\frac{1}{N_S^2} \sum_i^{N_S}\sum_{j\neq i}^{N_S}\mid\mid P^{QC}_{ij} - P^{ML}_{ij} \cdot p_i^{k} \cdot p_j^{k} \mid\mid^2 &\text{if}~dim(\mathbf{P})\leq 2 \end{array}
 \color{white}\right\}
\end{equation}{}
for vectorial and non-vectorial properties, respectively. The error $\varepsilon^{k}_P$ for a specific phase is computed as the mean squared error of a property $P$ from quantum chemistry (index $QC$) and machine learning (index $ML$). The property $P$ couples different states, indicated by $i$ and $j$. Since the wave function of each of the states can have an arbitrary phase, the property $P_{ij}$ that couples state $i$ and $j$ has to be multiplied with a product of the phases for these states, $p_i\cdot p_j$. The phases for all states together form a vector $p$ with entries of either +1 and -1. Which of the $2^{N_S-1}$ possible combinations for $p$ is chosen, is indicated by the index $k$, also defined in eq.~(\ref{eq:lph}). 
The relative signs within one vector remain and must be predicted correctly for successful training.

The overall loss function used in this work is a combination of such phase-less loss functions and mean squared errors obtained for all properties with a trade-off factor to account for their relative magnitude (a detailed description of the implementation is given in the SI).
This loss function removes the influence of the arbitrary phase during the learning process of a ML model and further reduces the computational costs for the training set generation.

\begin{figure}
    \centering
    \includegraphics[width=0.6\textwidth]{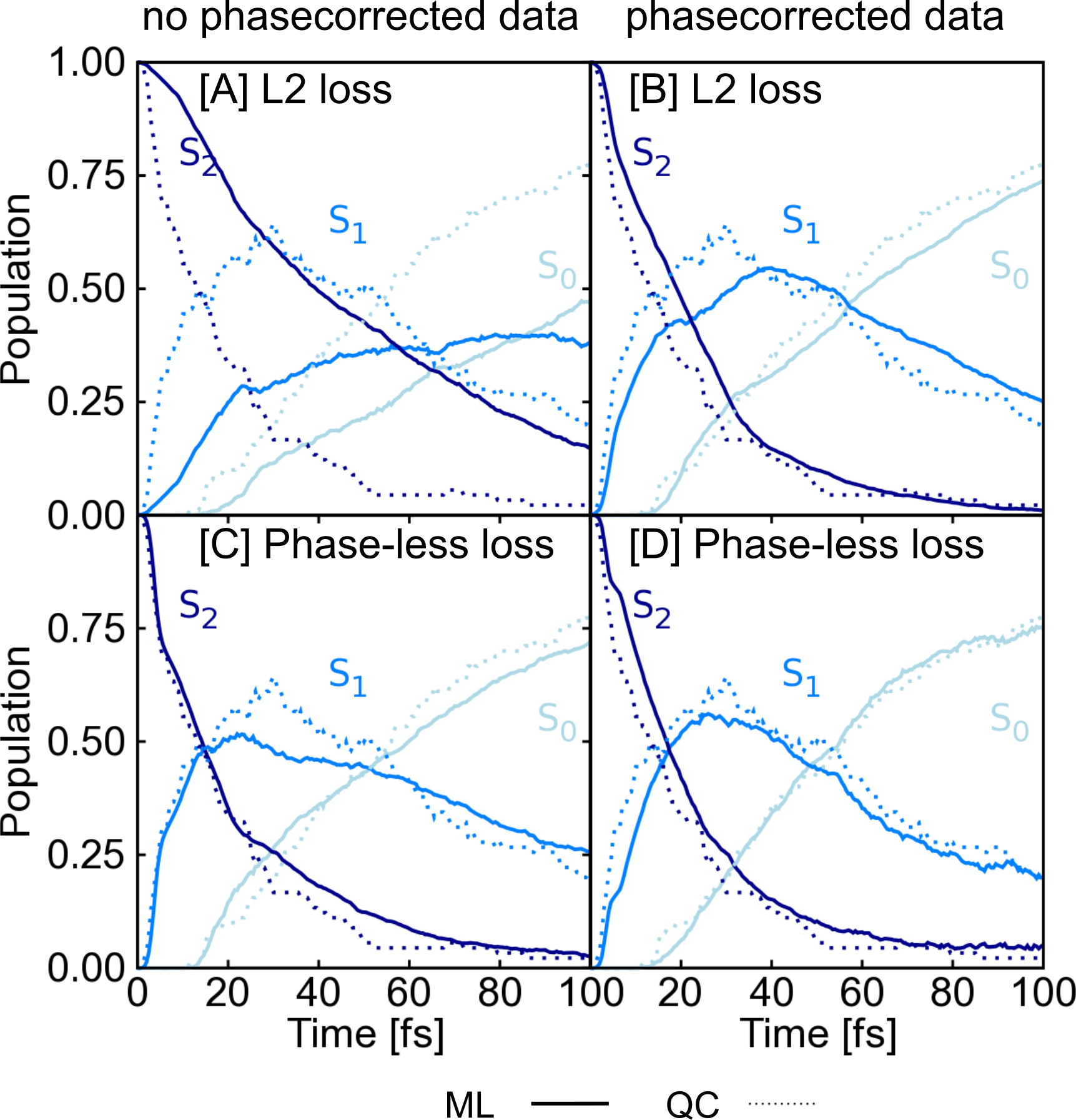}
    \caption{Populations obtained from 90 QC (MR-CISD(6,4)/aug-cc-pVDZ) trajectories are shown by continuous lines and are compared to (A) populations resulting from 1000 initially excited trajectories obtained from SchNet (dotted lines) that is trained on not phase corrected data and takes the L$_2$ norm as loss function, (B) a similar SchNet model, but trained on phase corrected data, (C) a SchNet model trained on not phase corrected data, but using the new phase-less loss function, and (D) a SchNet model trained on phase corrected data using the new phase-less loss function .} 
    \label{fig:phasecorrection}
\end{figure}

Results are given in Figure~\ref{fig:phasecorrection}, which shows the population schemes of CH$_2$NH$_2^+$ obtained after excitation to the second excited singlet state, S$_2$.
The SchNarc models are all trained on energies, forces, and NACs for the three considered singlet states.
The populations obtained from SchNarc models (panels A and C) trained on a data set that is not phase corrected, i.e. it contains couplings that can randomly switch their sign, are compared to populations obtained from models trained on phase corrected data (panels B and D). As can be seen, the L$_2$ loss function, as used in the upper plots, leads to an accurate ML model to reproduce ultrafast transitions only in the case of phase corrected data (panel C), whereas this loss can not be used when trained on raw quantum chemistry data (panel A). In comparison, a SchNarc simulation with a ML model that applies the phase-less loss function is successful in reproducing the populations for both training sets.

In those simulations, the NACs are multiplied with the corresponding energy gaps, i.e., $\widetilde{C}^{\text{NAC}}_{ij}=C^{\text{NAC}}_{ij}\cdot \Delta E_{ij}$, to get rid of singularities~\cite{Guan2019PCCP,Westermayr2019arXiv}. These smooth couplings $\widetilde{C}^{\text{NAC}}_{ij}$ are not directly learned, but rather constructed as the derivative of a virtual property, analogously to forces that are predicted as derivatives of an energy-ML model. The virtual property is the multi-dimensional anti-derivative of the rightmost expression in equation~(\ref{eq:nac5}), $\langle \Psi_i \mid \frac{\partial H_{el}}{\mathbf{R}}\mid\Psi_j\rangle$ (a derivation is given in the SI). Compared to previous ML models for NACs~\cite{Dral2018JPCL,Hu2018JPCL,Westermayr2019CS,Westermayr2019arXiv}, where NACs are learned and predicted as direct outputs or even single values, this approach provides rotational and translational covariance, which has recently been achieved in a similar way for the electronic friction tensor~\cite{Zhang2020JPCC}.

However, even without the need of pre-processing the training set, the costly computations of NAC vectors for the training set generation remain. 
Approximations of NACs exist and often involve the computation of the squared energy-gap Hessian~\cite{Thiel1999JCP,Koeppel2001JCP,Koeppel2006MP,Maeda2010JCTC,Kammeraad2016JPCL,Gonon2017JCP}.
Their use in dynamics simulations is rather impracticable with quantum chemistry methods, especially in the case of complex systems, due to the expenses of computing second-order derivatives.

Here, we take advantage of the efficiency for second-order derivative computation from ML models with respect to atomic coordinates to obtain the Hessians of the fitted PESs:

\begin{equation}
    \label{eq:hopdirML}
     \frac{\partial^2 (\Delta E_{ij})^2}{\partial R^2} = \frac{1}{2}\left(\Delta E_{ij} \cdot \frac{\partial^2 \Delta E_{ij}}{\partial R^2} + \left(\frac{\partial \Delta E_{ij}}{\partial R}\right)^2 \right) 
\end{equation}
with $R$ being the atomic coordinates of a molecular system. Note that Hessians are also employed in quantum dynamics simulations~\cite{Frankcombe2014JCP,Richings2015IRPC}, which might open further applications for our implementation. 

The squared energy-gap Hessian can be further obtained as the sum of two symmetric dyads, that define the branching space~\cite{Gonon2017JCP}. Hence, this Hessian can be employed to obtain the symmetric dyad of the smooth NACs via~\cite{Koeppel2001JCP,An2018CPL}:
\begin{equation}
     \widetilde{C}^{\text{NAC}}_{ij} \otimes \widetilde{C}^{\text{NAC}}_{ij} \approx \frac{\partial^2 (\Delta E_{ij})^2}{2\partial R^2} - \frac{\partial \Delta E_{ij}}{\partial R}\otimes\frac{\partial\Delta E_{ij}}{\partial R}.
\end{equation}{}
After singular value decomposition, the hopping direction can be computed as the eigenvector, $v_{ij}$, of the largest non-zero eigenvalue~\cite{Gonon2017JCP,Baeck2017JCP,An2018CPL} with the corresponding eigenvalue, $\lambda_{ij}$, as the squared magnitude of the ML smooth coupling, $\widetilde{C}^{\text{NAC}}_{ij}$. 
The final approximated NAC vectors, $C^{\text{aNAC}}_{ij}$, between two states are then:
\begin{equation}
 \label{eq:aNAC}
    C^{\text{aNAC}}_{ij}= v_{ij} \cdot \frac{\sqrt{\lambda_{ij}}}{\Delta E_{ij}}.
    \end{equation}{}

The approximated NAC vectors can be employed in the vicinity of a conical intersection, otherwise the output becomes too noisy. For the latter reason, we define thresholds of 0.5 eV and 1.0 eV for the energy gaps to compute approximated NACs between coupled singlet-singlet states and triplet-triplet states, respectively.
It is worth mentioning that the ML models slightly overestimate the energy gaps~\cite{Westermayr2019CS}, since, in contrast to quantum chemistry PESs, the ML PESs are smooth everywhere. 
In Ref.~\cite{Baeck2017JCP}, approximated NACs were applied for a 1D system and their usefulness in combination with ML was already anticipated. 

We turn this idea into reality and show ML excited-state dynamics with approximated NACs for a linear vibronic coupling (LVC) model~\cite{Plasser2019PCCP,Koeppel84ACP} of sulfur dioxide, SO$_2$, which we use as a reference system to assess the quality of approximated NACs and also include triplet states to treat SOCs. The LVC model of sulfur dioxide, $SO_2$,\cite{Plasser2019PCCP} contains 3 singlet states and 3 triplet states, with symmetry allowed NACs between the first and second excited singlet states as well as the first and third triplet states, and is used to train two ML models: One, where the NACs are learned and another, where the NACs are approximated according to equation~(\ref{eq:aNAC}). The ML models for dynamics simulations with approximated (trained) NACs are built up from energies, gradients, and SOCs (plus NACs) using 5,000 (20,000 for singlet only or 200,000 for singlet and triplet states) randomly selected data points from training sets consisting of 280,200 data points. 

In addition, we also apply the NAC approximation to two systems, where the data is obtained directly with ab initio methods: the methylenimmonium cation, CH$_2$NH$_2^+$, as presented before, and thioformaldehyde, CSH$_2$. 
CSH$_2$, is used, because it shows slow population transfer~\cite{Mai2019JCTC}, in contrast to the fast population transfer in CH$_2$NH$_2^+$. The training set is built up of 4,703 data points with two singlet states and two triplet states after initial sampling of normal modes and adaptive sampling with simple multi-layer feed-forward neural networks as done in Ref.~\cite{Westermayr2019CS} for CH$_2$NH$_2^+$. The program MOLPRO~\cite{MOLPRO} is used for these calculations with CASSCF(6,5)/def2-SVP. The dynamics simulations are carried out after excitation to the first excited singlet state for 3000 fs. A detailed analysis on the reference computations as well as information on the timing of the Hessian evaluation is given in the SI in sections S2 and S3.1 respectively.
None of the data points from the ab initio MD simulations, to which we compare our SchNarc models, are included in the training sets and, thus, the dynamics simulations can be seen as a direct test.

\begin{figure}
    \centering
    \includegraphics[width=0.8\textwidth]{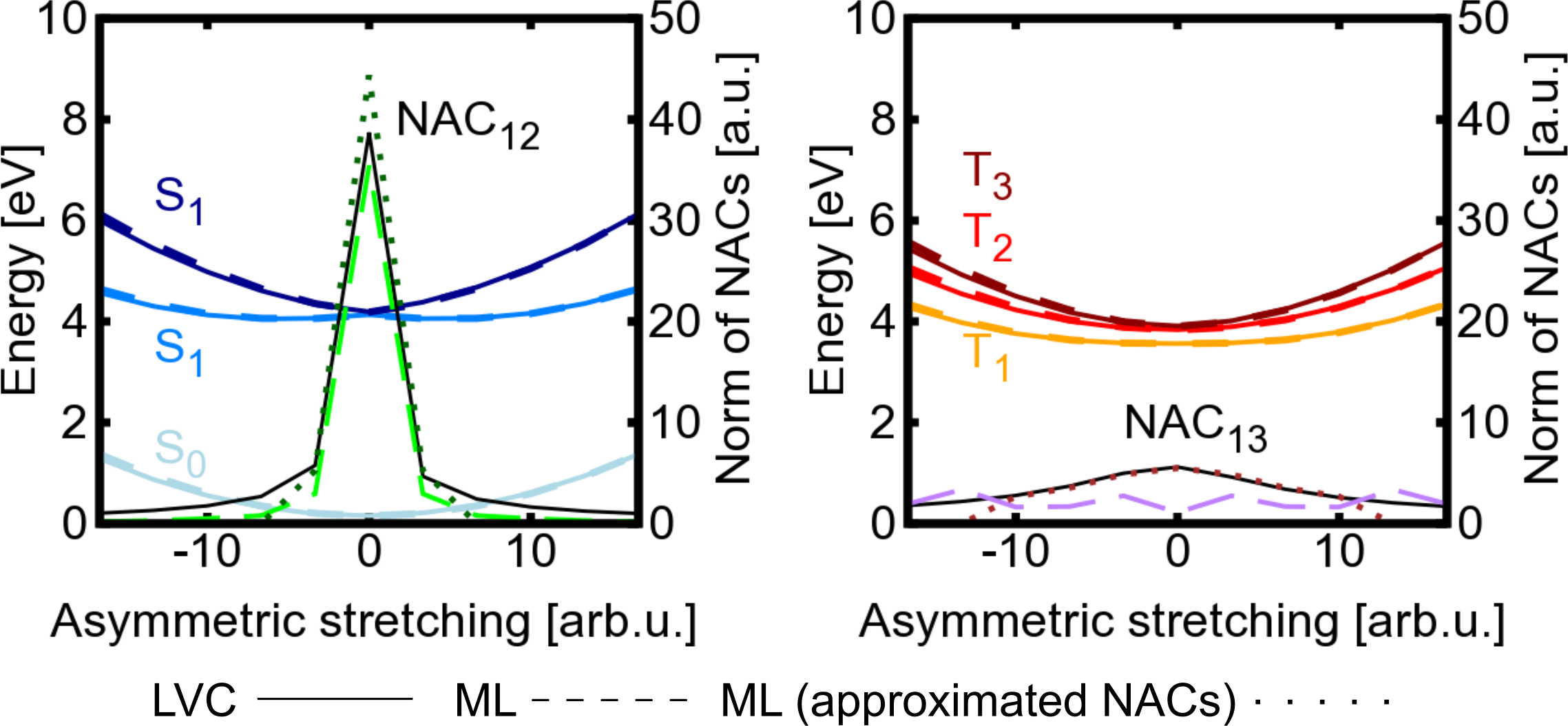}
    \caption{Potential energy curves and the norm of the NAC vectors between singlet states (A) and triplet states (B) along the asymmetric stretching mode of SO$_2$. Continuous line represent LVC(MR-CISD) and dotted (dashed) lines show results obtained from SchNarc models trained on only energies and gradients (as well as NACs for comparison) of 3 singlets and 3 triplet states.}
    \label{fig:PEC_SO2}
\end{figure}

Results for the LVC model are depicted in Fig.~\ref{fig:PEC_SO2}. The potential energy curves along the asymmetric stretching mode of the singlet states (left plot) and the triplet states (right plot) are shown along with the norm of the respective NAC vectors. As can be seen, the shape as well as the height of the peak of the norm of trained NACs (dashed lines) and approximated NACs (dotted lines) are comparable to those of the LVC model (continuous lines). The approximated NACs approach zero faster than the trained NACs, which is due to the applied threshold that is set for the energy gap to compute NACs. Noticeably, the learned NACs between the triplet states show a decrease, when the corresponding triplet energies are close to each other. This might be an effect due to the Berry phase, that can not be captured with our approach leading to artifacts in some regions.

\begin{figure}
    \centering
    \includegraphics[width=0.8\textwidth]{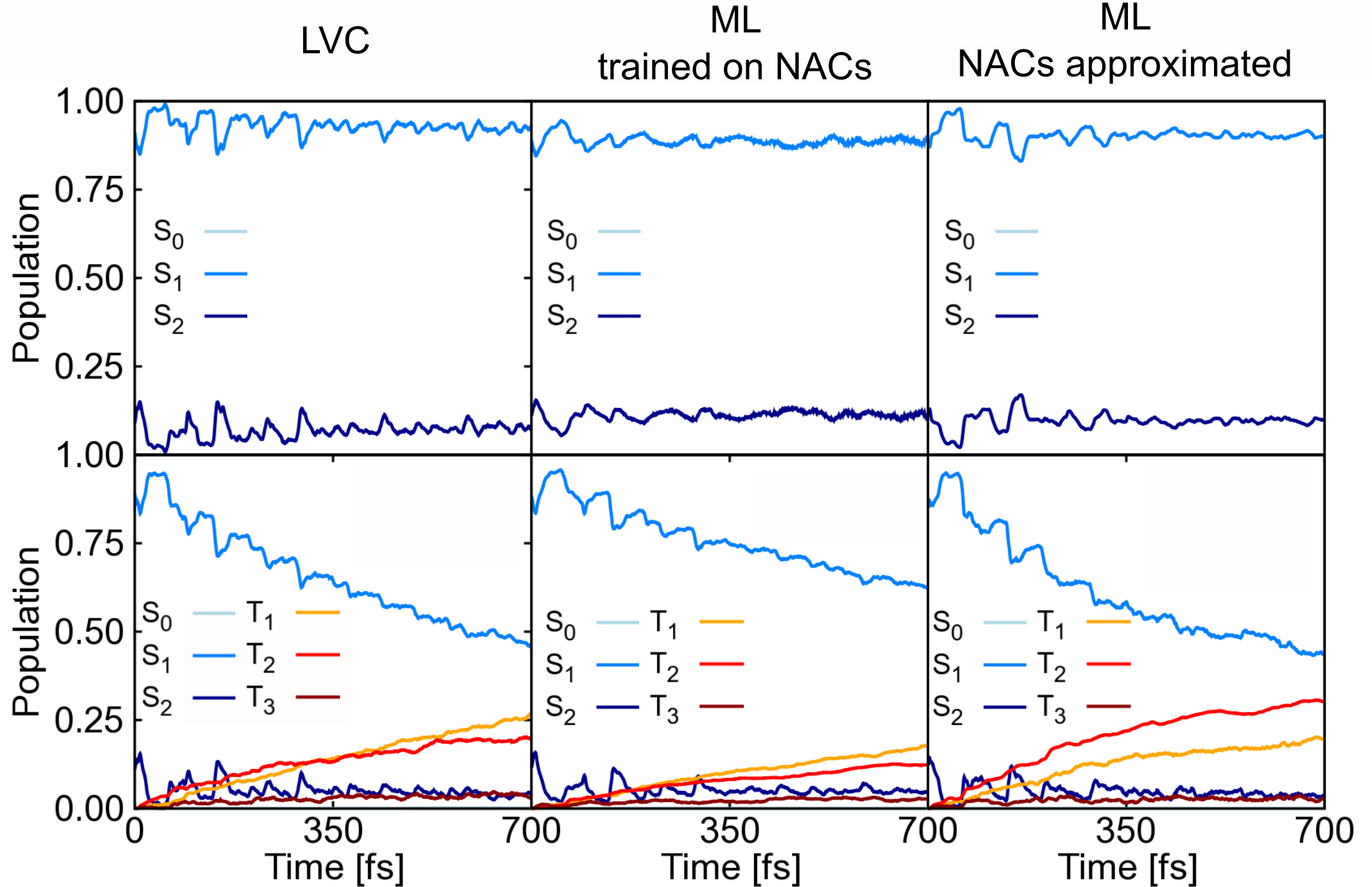}
    \caption{Quantum populations using LVC(MR-CISD) (left panels), SchNarc trained on NACs (middle panels) and SchNarc using approximated NACs from energies and gradients (right panels). Dynamics are shown of 1000 initially excited configurations considering only singlet states (upper line) and additionally triplet states (lower line).}
    \label{fig:popSO2}
\end{figure}

Importantly, the NACs can be predicted accurately with ML in most of the regions around the conical intersections. We first consider a singlet-only model in order to support this assumption with dynamics simulations, see Fig.~\ref{fig:popSO2}. LVC populations show minor population transfer between the second excited singlet state and the first excited singlet state, which can be reproduced with both SchNarc models (upper panels). 

Also when including triplet states, the SchNarc models can reproduce the dynamics (lower panels). Here, population is mostly transferred from the first excited singlet state to the triplet states. Note that populations in surface hopping are often only accurate to within 10\%, such that we judge the deviations of the ML populations from the LVC reference as small (see Ref.~\citenum{Westermayr2019arXiv} for examples, where dynamics is not reproduced by ML although potentials seemingly are). 

The application of the NAC approximation is further tested on more realistic systems with properties obtained from quantum chemistry data including fast as well as slow population transfer. CH$_2$NH$_2^+$ serves as a testsystem for the former case, where ultrafast transitions from the second excited singlet state back to the ground state take place after excitation within 100 femtoseconds~\cite{Barbatti2006MP,Westermayr2019CS}, which  is only possible to reproduce with accurate NACs~\cite{Westermayr2019arXiv}. In contrast, the CSH$_2$ molecule serves as a testsystem for slow populations transfer and shows intersystem crossing strongly dependent on the accuracy of the underlying potentials~\cite{Mai2019JCTC}. Inaccurate ML models would thus be unable to reproduce the reference dynamics.

\begin{figure}
    \centering
    \includegraphics[width=0.5\textwidth]{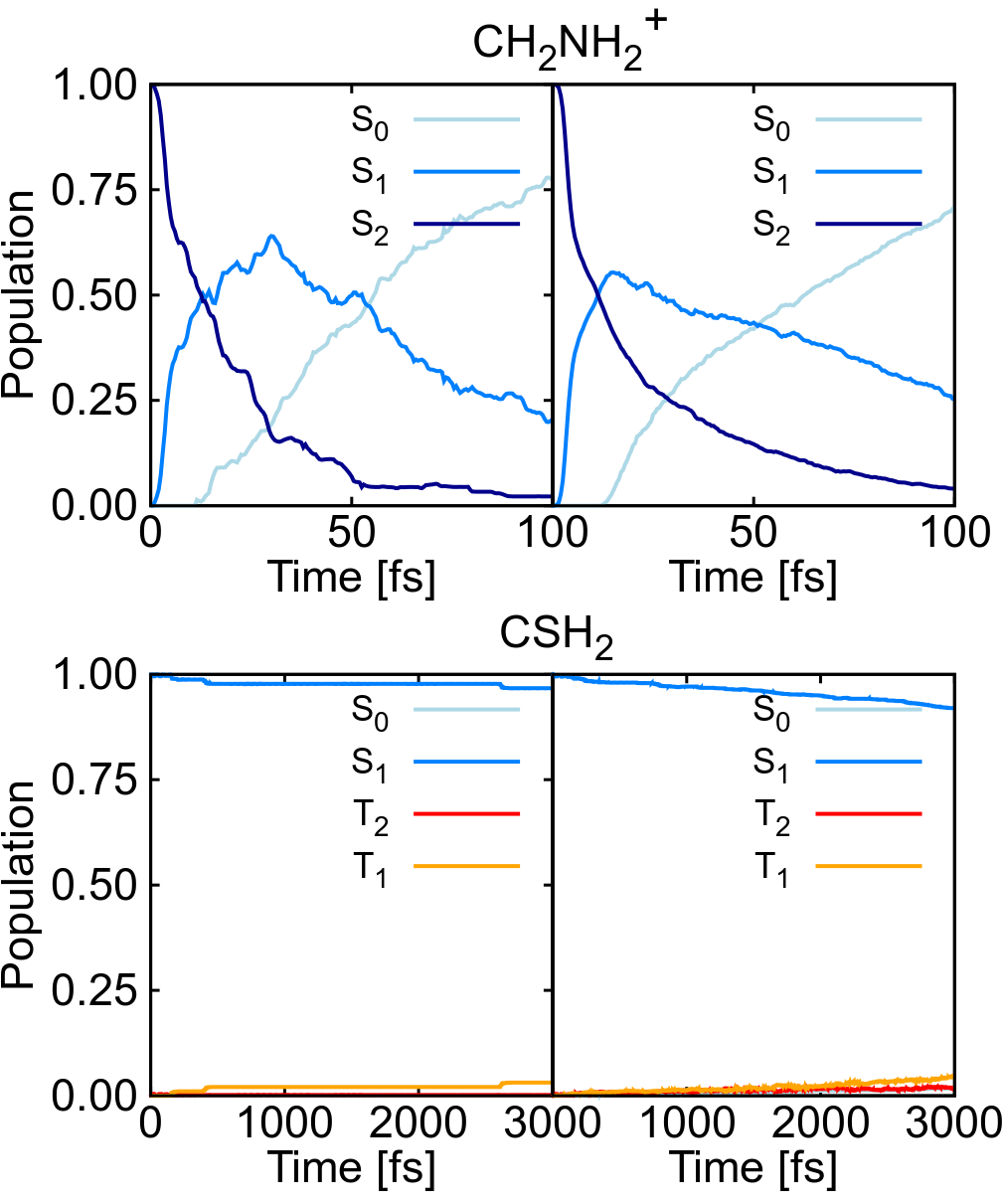}
    \caption{Quantum populations of the methylenimmonium cation obtained from (A) 90 trajectories using MR-CISD/aug-cc-pVDZ (QC) and (B) 1000 trajectories using SchNarc (ML) as well as populations up to 3,000 fs of the thioformaldehyde molecule obtained from (C) 100 trajectories using CASSCF(6,5)/def2-SVP (QC) and (D) 9590 trajectories using SchNarc (ML).}
    \label{fig:pop_molec}
\end{figure}

The populations of both systems are given in Fig.~\ref{fig:pop_molec}. The reference populations (left panels) are compared to SchNarc simulations with ML models trained on energies and gradients (panel B) as well as SOCs (panel D) for CH$_2$NH$_2^+$ and CSH$_2$, respectively. As can be seen, the fast and the slow population transfer can be reproduced accurately, which proves the validity of our ML approach.

In summary, the SchNarc framework combines the SHARC~\cite{Mai2018WCMS} approach for surface hopping and the SchNet~\cite{Schuett2019JCTC} approach for ML. The training of ML models is facilitated by using the phase-less loss and the NAC approximation, avoiding quantum chemical NAC calculations at all. Thus, photodynamics simulations are possible solely based on ML PESs, their derivatives and SOCs. Furthermore, this method allows for a very efficient computation of the Hessians of all the excited states at each time step. Hence, SchNarc allows for efficient nonadiabatic dynamics simulations of excited states and light-induced processes including internal conversion and intersystem crossing.

\section*{Acknowledgement}
This  work was financially supported by the AustrianScience Fund, W 1232 (MolTag), the uni:docs program of the University of Vienna (J.W.), the HPC-Europa3 program (J.W.), and the European Union Horizon 2020 research and innovation program under the Marie Sklodowska-Curie grant agreement NO 792572 (M.G.). The computational results presented have been achieved in part using the Vienna Scientific Cluster (VSC). All authors thank Univ.-Prof. Dr. Klaus-Robert M\"{u}ller for hosting the visit of J.W. in the frame of HPC-Europa 3. P. M. thanks the University of Vienna for continuous support, also in the frame of the research platform ViRAPID. J.W. and P. M. are grateful for an NVIDIA Hardware Grant 
and thank Sebastian Mai for helpful discussions concerning the phase-free training algorithm.

\section*{Data availability}

The data sets are partly available~\cite{Westermayr2019CS} and will be partly made available on github.com/schnarc. The molecular geometries and corresponding properties are saved in a database format provided by the atomic simulation environment~\cite{Larsen2017IOPP}.

\section*{Code availability}
All code developed in this work will be made available on github.com/schnarc.

\section*{Supplementary Information}

\section{SchNet for excited states}
As a machine learning (ML) model, the deep continuous-filter convolutional neural network SchNet, that is described in detail in Ref.~\cite{Schuett2018JCP,Schuett2019JCTC} is used and adapted for excited states to train excited-state energies, forces, spin-orbit couplings (SOCs), and nonadiabatic couplings (NACs). 

The molecular descriptor is constructed by SchNet~\cite{Schuett2018JCP} that treats atoms in their chemical and structural environment. A cutoff is defined to specify the environment that is included for the description of an atom. Hence the molecular properties are obtained as atom-wise contributions.
A continuous-filter convolutional layer and several additional interaction layers define and optimize the atom representations. These representations are mapped to different properties via fully connected layers with shifted softplus activation functions. These prediction blocks, which use a common descriptor network, are separately designed for energies, SOCs, and NACs, whereas the forces are derived with respect to atomic coordinates from outputs of the ML model for energies. The loss function is a combined loss function of all the properties. A trade-off is defined to weigh the properties according to their magnitude. The properties that should be learned have to be specified along with the corresponding trade-off in an additional input file.

\subsection{Standard loss function, $L_2$}
The overall $L_2$ loss function as implemented in SchNet for excited states, reads:

\begin{equation}\label{eq:l2}
\begin{array}{ll}
      L_2 = t_E\mid\mid E^{QC} - E^{ML} \mid\mid^2 + t_F \mid\mid F^{QC}-F^{ML}\mid\mid^2 + \\
     t_{\text{SOC}} \mid\mid C_{\text{SOC}}^{QC}-C_{\text{SOC}}^{ML}\mid\mid ^2 + t_{\text{NAC}}\mid\mid C_{\text{NAC}}^{QC}-C_{\text{NAC}}^{ML}\mid\mid ^2 , 
\end{array}{}
\end{equation}{}
where $t_E, t_F, t_{SOC},$ and $t_{NAC}$ define the trade-offs for the properties E (energies), F (forces), $C_{\text{SOC}}$ (SOCs), and $C_{\text{NAC}}$ (NACs), respectively. Corresponding labels with an index "QC" refer to the reference value and with an index "ML" to the the SchNet predictions. 

\subsection{Phase-less loss function, $L_{ph}$}
In order to train on inconsistent SOCs and NACs with respect to their sign, we have developed a phase-less loss-function. This is based on the $L_2$ loss, but here, the squared error of the predicted properties, $P$, is computed more often, i.e. $2^{N_S-1}$-times with $N_S$ being the total number states. The value, $L_P$, that enters the overall loss function, $L_{ph}$, is the minimum function of all possible squared errors, $\varepsilon^k_{P}$, for a given property, $P$:

\begin{equation}
    \label{eq:lphSI}
    L_P = \text{min}\left(\{\varepsilon^{k}_{P}\}\right) ~~ \text{with} ~ 0 \leq k \leq 2^{N_S-1}
    \end{equation}
    with 
    \begin{equation}
      \varepsilon^{k}_{P} =  \left\{
\begin{array}{ll}
 \sum_i^{N_S}\sum_{j\neq i}^{N_S}\frac{1}{N_A} \sum_m^{N_A} \mid\mid P^{QC}_{ij,m} - P^{ML}_{ij,m} \cdot p_i^{k} \cdot p_j^{k} \mid\mid^2  & \text{if}~dim(\mathbf{P})\geq 3\\
 \sum_i^{N_S}\sum_{j\neq i}^{N_S}\mid\mid P^{QC}_{ij} - P^{ML}_{ij} \cdot p_i^{k} \cdot p_j^{k} \mid\mid^2 &\text{if}~dim(\mathbf{P})\leq 2 \end{array}
 \color{white}\right\}
\end{equation}{}
for vectorial and non-vectorial properties, respectively. The error $\varepsilon^k_P$ for a specific phase is computed as the mean squared error of a property $P$ from quantum chemistry (index $QC$) and machine learning (index $ML$). The property $P$ couples different states, indicated by $i$ and $j$. Since the wave function of each of the states can have an arbitrary phase, the property $P_{ij}$ that couples state $i$ and $j$ has to be multiplied with a product of the phases for these states, $p_i\cdot p_j$. The phases for all states together form a vector $p$ with entries of either +1 and -1. Which of the $2^{N_S-1}$ possible combinations for $p$ is chosen, is indicated by the index $k$, also defined in eq.~(\ref{eq:lphSI}). Since we are free to choose one of the phases, we set the phase of the first state always to +1. The relative signs within one vector remain and must be predicted correctly for successful training.

The overall loss function used in this work is a combination of such phase-less errors and mean squared errors obtained for all properties with a trade-off factor to account for their relative magnitude (as already specified in equation \ref{eq:l2}):

\begin{equation}
    L_{ph} = t_E\mid\mid E^{QC} - E^{ML} \mid\mid^2 + t_F \mid\mid F^{QC}-F^{ML}\mid\mid^2 + t_{SOC} \cdot L_{SOC} + t_{NAC} \cdot L_{NAC}
\end{equation}{}
This cost function removes the influence of the arbitrary phase during the learning process of a ML model and further reduces the computational costs for the training set generation.

We tested on several alternatives, such as a loss function that additionally includes the norm of a vector, variations of a minimum function and another type of phase-free loss function, that can be used if only one type of coupling, i.e. SOCs or NACs, or dipole moments are trained. This error is also implemented in SchNet for excited states and the error of a property that couples state i and j, $\varepsilon^k_{P}$ is computed as follows:

\begin{equation}
\label{eq:lph1}
\varepsilon^{k,\pm}_{P} = 
\left\{
\begin{array}{ll}
 \mid\mid P^{QC}_{ij} \pm P^{ML}_{ij}\mid\mid^2 &\text{if}~dim(\mathbf{P})\leq 2\\
 \frac{1}{N_A}\sum_m^{N_A} \mid\mid P^{QC}_{ij,m} \pm P^{ML}_{ij,m}\mid\mid^2 &\text{if}~dim(\mathbf{P}) \geq 3
\end{array}
\right.
\end{equation}
As can be seen, the error is computed twice -- once assuming a correct phase of predicted properties and once a phase switch, that are both combined subsequently in case $dim(\mathbf{P})\geq 3$:
\begin{equation}
    e_P^k = \varepsilon_p^{k,-}\cdot C_{ij}^+ + \varepsilon_p^{k,+}\cdot C_{ij}^-
\end{equation}{} with
\begin{equation}
    C_{ij}^{\pm} = \frac{\varepsilon_{p}^{k,\pm}}{\varepsilon_p^{k,-} + \varepsilon_{p}^{k,+}}.
\end{equation}{}
The value, $L_P$, that enters the loss function is then either a combination of both possibilities for vectorial properties or a minimum function of the two possible errors, $\varepsilon_P^{k,\pm}$.
This variation gives comparably accurate results and also leads to a phase-free training. Experiments have shown, that -- e.g. in the case of the SO$_2$ molecule -- more data points, but shorter training is necessary. This alternative error can be more favorable in cases, where only one coupling type is needed and many states are involved, since computation of all possible combinations can be omitted. 
All other tested variations turned out to be less successful in learning the shape of couplings.
\subsection{Machine learning models}
The model parameters for each molecular system are given in Table~\ref{tab:ML} including the number of data points and states trained. The errors for the remaining test set (i.e. the data points not used for training and validation) are listed as mean absolute values resulting from all states. For the CH$_2$NH$_2^+$ and CSH$_2$ models, the ML predictions reach chemical accuracy and in some cases the error is even below 0.043 eV (1 kcal/mol).  If not stated otherwise, 256 features with 3 hidden layers are used for each model. The batch size ranges from 20 to 50 and is set in order to comply with the maximum allowed memory of a used GPU. The learning rate is set to 0.0001 and is reduced by a factor of 0.8 down to a value of 0.000001 with a patience of 15 steps. The maximum number of epochs is set to 5000. The trade-off for each property is defined, so that the mean squared errors in the first few epochs is equally large for all properties. 
\begin{table}
\caption{\label{tab:ML}Parameters for the trained SchNet models for excited states. For SO$_2$, CSH$_2$, and CH$_2$NH$_2^+$ 200-5,000, 200 and 100 data points are used for validation, respectively, hence at least 50,000, 503, and 900 data points are used for testing. The MAEs and RMSEs are reported in eV, eV/\AA, and a.u. for energies, gradients and all type of couplings, respectively. If not mentioned otherwise, the data sets are not phase corrected.}
\begin{tabular}{c|c|c|c|c|c|c}
     Molecule & training  &S/T&properties (t) &cutoff&Loss&MAE (RMSE)\\
     &points &&& [\AA] && \\
     \hline 
     SO$_2$ &5,000 &3/0&E (1.0) &5.0 &L2& 0.069 (0.20) \\
      & & &F (0.25) &&& 0.20 (0.67)\\
     
     SO$_2$ &20,000&3/0&E (1.0)  &8 .0 &L$_{ph}$ &0.062 (0.184)\\
     & & &F (0.1) &&&0.24 (0.65)\\
     & & &NAC (0.004)&&&0.13 (1.15) \\

     SO$_2$ &5,000&3/3&E (1.0)  &5.0 &L$_{ph}$ &0.029 (0.068)\\
     & & &F (0.25) &&& 0.12 (0.26)\\
     & & &SOC (300) &&&$7.7\cdot 10^{-6}$ ($3.1\cdot 10^{-5}$)\\
     
     SO$_2$ &20,000&3/3&E (1.0)  &8.0 &L$_{ph}$ &0.027 (0.068)\\
     & & &F (0.25)&& &0.11 (0.26)\\
     & & &NAC (0.0001)&& &0.52 (23.8) \\
     & & &SOC (300) &&&$1.2\cdot 10^{-5}$ ($4.4\cdot 10^{-5}$)\\
     
     \hline

     CH$_2$NH$_2^+$ &3,000  &3/0 &E (1.0) &10.0 &L2 &0.059 (0.13)\\
     &&&F(1.0) & &&0.15 (0.30)\\
     & &&NAC (0.001) & &&0.22 (0.89)\\
          
     CH$_2$NH$_2^+$ &3,000  &3/0 &E (1.0) &10.0 &L$_{ph}$ &0.059 (0.14)\\
     & &&F(1.0) & & & 0.14 (0.32)\\
     & &&NAC (0.004) & && 0.15 (0.55)\\

     \hline
     CH$_2$NH$_2^+$ &3,000 &3/0 &E (1.0) &10.0 &L2& 0.042 (0.087)\\
      &\textit{phase} &&F(1.0) & && 0.096 (0.22)\\
     &\textit{corrected}&&NAC (0.0001) & &&0.21 (0.83)\\
     CH$_2$NH$_2^+$ &3,000 &3/0 &E (1.0) &10.0 &L$_{ph}$ &0.050 (0.16)\\
     &\textit{phase} &&F(1.0) & &&0.13 (0.32) \\
     &\textit{corrected}&&NAC (0.004) & && 0.15 (1.1)\\
         CH$_2$NH$_2^+$ &3,000 &3/0 &E (1.0) &10.0 &L2 &0.048 (0.12)\\
     &&&F(1.0) & & &0.13 (0.30)\\
     
     \hline     
     CSH$_2$ &4,000 &2/2 &E (1.0) &10.0 &L$_{ph}$ &$4.1\cdot10^{-4}$ ($6.1\cdot 10^{-4}$)\\
     &&&F(1.0) & & &$6.2\cdot10^{-4}$ ($1.1\cdot 10^{-3}$)\\
     &&&SOC (500) && &$6.1\cdot10^{-6}$ ($1.6\cdot 10^{-5}$)\\

\end{tabular}
\end{table}
\newpage
\section{Training sets and reference computations}
\subsection{Training set generation}
SchNet models for excited states are trained on a linear vibronic coupling model (LVC) of SO$_2$~\cite{Plasser2019PCCP,Koeppel84ACP}, the methylenimmonium cation, CH$_2$NH$_2^+$ and thioformaldehyde, CSH$_2$. 

The molecular geometries and corresponding properties are saved in a database format provided by the atomic simulation environment~\cite{Larsen2017IOPP}. No data points of the dynamics simulations to which we compare SchNarc models are included in the training sets. 
The phase corrected training set for the methylenimmonium cation, CH$_2$NH$_2^+$, is taken from Ref.~\cite{Westermayr2019CS} and consists of 4000 data points. The geometries of this training set were recomputed with the same level of theory, MR-CISD(6,4)/aug-cc-pVDZ, but without applying any pre-processing, such as phase correction, in order to provide a non-phase corrected training set. The program suite COLUMBUS~\cite{Lischka2001PCCP} was used for this purpose, resulting in 3998 converged single point calculations. For the dynamics simulations with SchNarc, 1000 trajectories (resulting from 20,000 initial conditions sampled from a Wigner distribution~\cite{Wigner1932PR}) are propagated for 100 fs using a time step of 0.5 fs. 

The training set for thioformaldehyde, CSH$_2$, is generated in the same way as it is done in Ref.~\cite{Westermayr2019CS} for CH$_2$NH$_2^+$. Initial configurations are sampled via scans of different reaction coordinates, such as normal modes. Additionally, adaptive sampling for excited states is carried out using two simple multi-layer feed-forward neural networks. At a number of 4855 data points, the networks seem to be converged and dynamics simulations up to 3 ps can be reproduced. The training set consists of 4703 data points, where samples showing a smaller energy gap than 0.01 H between triplet-triplet states are sorted out due to problematic data points in those regions. Without these points the NNs converge much better, in about half of the time. The RMSE of forces is slightly larger, the rest of the errors are comparable. The level of theory is CASSCF(6,5)/def2-SVP and 2 singlet and 2 triplet states are included. Quantum chemistry calculations are carried out using Molpro~\cite{MOLPRO}. In addition to energies, forces, SOCs, and NACs, the permanent and transition dipole moments are included in the training set.
For the dynamics simulations of CSH$_2$, 40,000 initial conditions are sampled from a Wigner distribution~\cite{Wigner1932PR} and excited to the first excited singlet state (2.0-2.5 eV). 100 trajectories are propagated with the reference method for 3 ps with a time step of 0.5 fs. The resulting populations are compared to 959 
trajectories obtained from SchNarc. 

For the SO$_2$ molecule, we refer to dynamics simulations with the reference method to generate the training set, which is based on a "one-shot" LVC model~\cite{Plasser2019PCCP}. 10,000 initial conditions are sampled from a Wigner distribution~\cite{Wigner1932PR} and excited between 0 and 10~eV. Surface hopping molecular dynamics simulations are carried out with SHARC after excitation of 1200 initially sampled geometries. They are propagated with NAC vectors for 700 fs with a time step of 0.5 fs. The first 200 trajectories are taken for the training set generation. This procedure is done twice - once only singlet states are considered and once singlet and triplet states are taken into account, resulting in 280,200 data points for each training set. The remaining 1000 trajectories serve for comparison to SchNarc dynamics. 
Due to symmetry, the SO$_2$ model contains NACs only between the S$_1$ state and the  S$_2$ state as well as between the T$_1$ state and the T$_3$ state. We considered this restriction for the SchNarc computations by setting the other couplings to zero.

It is worth mentioning that SO2 needs more data points for training than CSH$_2$ and CH$_2$NH$_2^+$,  since for the latter molecules, adaptive sampling was applied and for SO$_2$ we used data directly from dynamics simulations with the LVC model. The dynamics with the LVC model are extremely fast and hence for the training set generation the usual adaptive sampling approach~\cite{Behler2015IJQC,Gastegger2017CS,Westermayr2019CS} is far more costly and time-intensive. Since reference computations are even cheaper than ML predictions, it is not our goal to provide a perfect training set with a minimum number of data points for this model, but rather to provide an easy-to-use but yet challenging test system to validate our method.

\subsection{Surface hopping molecular dynamics}
In order to compute nonadiabatic molecular dynamics simulations, the SHARC~\cite{Mai2018WCMS,Richter2011JCTC,sharc-md2} method, an extension of Tully's fewest switches algorithm~\cite{Tully1990JCP}, is applied. This mixed quantum-classical approach allows for on-the-fly computation of the PESs with electronic structure methods, on which the nuclei move according to Newton's second equation of motion. In order to account for nonadiabatic transitions between states of same spin multiplicity, instantaneous switches from one state to the others are allowed in regions of high hopping probability. 
After every simulation, the trajectories are analyzed using the SHARC diagnostic tools to check improper behaviour, such as energy fluctuations. A few reference trajectories are sorted out in each case -- mainly due to improper convergence of quantum chemistry calculations in critical regions of the potential energy surfaces. Decoherence correction is applied~\cite{Granucci2010JCP} and the hopping probabilities are computed from SOCs and NACs from electronic structure calculations in case of reference dynamics or from ML models in case of SchNarc dynamics~\cite{sharc-md2}. The velocities are corrected along the direction of the NAC vectors in each simulation and for dynamics with quantum chemistry, the phase is tracked along an independent trajectory.

\section{Nonadiabatic couplings}
As mentioned in the main text, NACs are either approximated or derived from a virtual property built by SchNarc. 
In order to define the virtual property, which SchNarc builds internally, we start by the derivative of the electronic Hamiltonian, $H_{el}$, with respect to the atomic coordinates of a molecule, $\mathbf{R}$:

\begin{eqnarray}
\label{eq:nac2}
\nonumber
\frac{\partial {H_{el_{ij}}(r,\mathbf{R})}}{\partial \mathbf{R}} \\
=\frac{\partial}{\partial \mathbf{R}} \langle \Psi_i \mid {H_{el}(r,\mathbf{R})} \mid \Psi_j \rangle \\
\nonumber
= \langle \frac{\partial}{\partial \mathbf{R}}\Psi_i \mid H_{el}(r,\mathbf{R}) \mid \Psi_j \rangle
    + \langle \Psi_i \mid \frac{\partial H_{el}(r,\mathbf{R})}{\partial \mathbf{R}} \mid \Psi_j \rangle 
    + \langle \Psi_i \mid H_{el}(r,\mathbf{R}) \mid  \frac{\partial}{\partial \mathbf{R}}\Psi_j \rangle
\end{eqnarray}

Since the adiabatic wavefunctions are eigenfunctions of $H_{el}(r,\mathbf{R})$, we can reformulate equation (\ref{eq:nac2}):
\begin{eqnarray}
    \label{eq:nac3}
    \nonumber
    \frac{\partial {H_{el_{ij}}(r,\mathbf{R})}}{\partial \mathbf{R}} \\
    = E_j \langle \frac{\partial}{\partial \mathbf{R}} \Psi_i \mid \Psi_j \rangle+ \langle\Psi_i  \mid \frac{\partial H_{el}(r,\mathbf{R})}{\partial \mathbf{R}} \mid \Psi_j \rangle + E_i \langle \Psi_i \mid \frac{\partial }{\partial \mathbf{R}} \Psi_j \rangle
\end{eqnarray}

By using the relation, $\langle \Psi_i \mid \frac{\partial}{\partial \mathbf{R}}\Psi_j \rangle = - \langle \frac{\partial}{\partial \mathbf{R}} \Psi_i \mid \Psi_j \rangle$, we can write:

\begin{equation}
    \label{eq:nac4}
    \frac{\partial H_{el_{ij}}(r,\mathbf{R})}{\partial \mathbf{R}} = (E_i-E_j) \cdot \langle  \Psi_i \mid \frac{\partial}{\partial \mathbf{R}} \Psi_j \rangle + \langle \Psi_i \mid \frac{\partial H_{el}}{\partial \mathbf{R}} \mid \Psi_j \rangle.
\end{equation}{}

Applying the Hellmann-Feynman theorem~\cite{Feynman1939PR}, we obtain the diagonal elements as the gradients,
\begin{equation}
    \label{eq:gradients}
    \frac{\partial}{\partial \mathbf{R}} H_{el_{ij}}= \nabla E_{ij} ~~~for ~ i = j,
\end{equation}
and obtain the NAC terms as properties that are inversely proportional to the corresponding energy gap of two adiabatic electronic states:
\begin{equation}
    \label{eq:nac5SI}
        C^{\text{NAC}}_{ij} \approx \langle  \Psi_i \mid \frac{\partial}{\partial \mathbf{R}}\Psi_j \rangle = \frac{1}{E_i-E_j}\langle \Psi_i \mid \frac{\partial H_{el}}{\partial \mathbf{R}}\mid \Psi_j \rangle    ~~~\text{for}~  i \neq j.
\end{equation}

The virtual property that SchNarc is generating is then the multi-dimensional anti-derivative of the latter expression in equation~\ref{eq:nac5SI}, $\langle \Psi_i \mid \frac{\partial H_{el}}{\mathbf{R}}\mid\Psi_j\rangle$. 
Noticeably, 
due to the Berry phase~\cite{Herzberg1963DFS,Worth2004ARPC,Ryabinkin2017ACR} the NAC vector field is not conservative~\cite{Koeppel2001JCP} and a line integral remains path dependent. Hence this approach does not include the effects of the Berry phase, which is also neglected in approaches such as the Zhu-Nakamura approximation~\cite{Oloyede2006JCP,Ishida2017IRPC} that does not contain a phase at all, or the phase correction algorithm~\cite{Akimov2018JPCL,Westermayr2019CS}. The mixed ML-classical dynamics are thus assumed to be mostly unaffected~\cite{Akimov2018JPCL,Westermayr2019CS}, which might not be the case in quantum dynamics simulations.

\subsection{NAC approximation and timing}
The approximation of NAC vectors, as explained in the main text and adapted from Refs.~\cite{An2018CPL,Baeck2017JCP,Koeppel2001JCP}, relies on the approximation of the Hessian from energy potentials between two states. It is especially powerful for ML models trained on quantum chemistry methods, where implementations of NAC vectors are largely missing, such as linear-response methods and here especially between the first excited state and the ground state, with the ADC(2) method being a prominent example~\cite{Dreuw2015WCMS}. Such an ML approach could further be used to pave the way towards efficient Hessian computations for all the states treated in quantum dynamics simulations using the variational multi-configurational Gaussian method, where the direct Hessian computation often remains the time limiting step~\cite{Richings2015TF,Frankcombe2014JCP}.

The used NAC approximation is valid in the vicinity of a conical intersection and hence relies on a threshold to define the energy gap, for which the approximation is applied.  In order to avoid additional computations, the Hessians are thus only computed if one of the energy differences between all possible singlet-singlet or triplet-triplet potentials is within the given threshold (as default we set 0.5 eV and 1.0 eV for singlet-singlet and triplet-triplet gaps). Since the gaps of the PESs are overestimated in case of CH$_2$NH$_2^+$~\cite{Westermayr2019CS}, this threshold is increased by 30\%. This means, that based on this pre-defined threshold, SchNarc decides whether NAC vectors are computed or not. In all other cases, the NAC vectors are set to zero. It is advisable to check the used thresholds for certain cases and adapt them, where necessary. It is worth mentioning, that this approach is limited to same-symmetry electronic states, which are, nevertheless, the most probable avoided state crossings in case of real, polyatomic systems~\cite{Baeck2017JCP}. 

\subsection*{Timing}
For the thioformaldehyde molecule, the evaluation of the 4 Hessians takes approximately 2 seconds, for the methylenimmonium cation, the computation of 3 Hessians takes 3-4 seconds, both on a CPU. A test computation of a 24 atom molecule was further carried out with SchNarc, which showed that the evaluation of 1 Hessian took around 45 sec on a CPU, which could be reduced to 14 sec on a GPU. In future work, we thus seek to adapt the code in order to compute only the relevant Hessians, i.e., those of close-lying states with respect to the active state during a dynamics simulation. To this aim, we seek to give the information of the active state to the SchNarc model, which is not yet implemented for our pySHARC~\cite{Plasser2019PCCP,Westermayr2019CS} wrapper (python wrapper for the SHARC code, which avoids heavy file I/O). Hence the Hessians would only be computed for the states that are close enough to the active state and the dynamics simulations are then still very efficient compared to pure quantum chemistry dynamics simulations. A comparison of the timings of 100 time steps for a dynamics simulation with the current SchNarc implementations and SHARC is given in Table~\ref{tab:my_label}. 

As can be seen, the LVC dynamics using the pySHARC wrapper are very cheap and only serve as a test system. It is however clearly visible that the training of NACs in this case needs way more data points and hence it takes longer to train the models. The dynamics of the CH$_2$NH$_2^+$ molecule using the MR-CISD method are expensive and ML can substantially decrease the simulation time. It can be seen that the Hessian computation with ML becomes more expensive, the larger the molecule becomes or the more states are involved in a simulation.

\begin{table}[]
 \caption{Comparison of the timings of 100 steps of a dynamics simulations using SchNarc with learned and approximated NACs as well as SHARC with quantum chemistry or the LVC model for SO$_2$. }
    \centering
    \begin{tabular}{c|c|c|c|c }
         &\# &\multicolumn{3}{c}{{100 time steps [s/CPU]} \textit{(Training[h/GPU]/data points)}}\\
         &States&\textbf{SchNarc} &\textbf{SchNarc} &\textbf{SHARC} \\
          & S/T& NAC learned &NAC excluded\\
         \hline
         SO$_2$&3/0 &5 (\textit{50.8/20,000}) &6 (\textit{3.7/5,000}) & 1-2\\
         SO$_2$&3/3 &13 (\textit{308/200,000}) &17(\textit{18.3/5,000}) & 1-2\\
         CSH$_2$ &2/2 &7 (\textit{13.0/4,000}) &8 (\textit{12.9/4,000})&52\\
         CH$_2$NH$_2^+$ &3/0 &12 (\textit{19.9/3,000})&126 (\textit{11.1/3,000}) & 37,112\\
    \end{tabular}
   
    \label{tab:my_label}
\end{table}{}


\begin{thebibliography}{100}

\bibitem{Cerullo2002S}
G.~Cerullo, D.~Polli, G.~Lanzani, S.~De~Silvestri, H.~Hashimoto, R.~J. Cogdell,
  Photosynthetic Light Harvesting by Carotenoids: Detection of an Intermediate
  Excited State, \emph{Science}, \textbf{298}, 2395--2398 (2002).

\bibitem{Schultz2004S}
T.~Schultz, E.~Samoylova, W.~Radloff, I.~V. Hertel, A.~L. Sobolewski,
  W.~Domcke, Efficient Deactivation of a Model Base Pair via Excited-State
  Hydrogen Transfer, \emph{Science}, \textbf{306}, 1765--1768 (2004).

\bibitem{Schreier2007S}
W.~J. Schreier, T.~E. Schrader, F.~O. Koller, P.~Gilch, C.~E.
  Crespo-Hern\'{a}ndez, V.~N. Swaminathan, T.~Charell, W.~Zinth, B.~Kohler,
  Thymine Dimerization in DNA Is an Ultrafast Photoreaction, \emph{Science},
  \textbf{315}, 625--629 (2007).

\bibitem{Rauer2016JACS}
C.~Rauer, J.~J. Nogueira, P.~Marquetand, L.~Gonz\'alez, Cyclobutane Thymine
  Photodimerization Mechanism Revealed by Nonadiabatic Molecular Dynamics,
  \emph{J. Am. Chem. Soc.}, \textbf{138}, 15911--15916 (2016).

\bibitem{Romero2017N}
E.~Romero, V.~I. Novoderezhkin, R.~v. Grondelle, Quantum design of
  photosynthesis for bio-inspired solar-energy conversion, \emph{Nature},
  \textbf{543}, 355 (2017).

\bibitem{mathew2014NC}
S.~Mathew, A.~Yella, P.~Gao, R.~Humphry-Baker, B.~F.~E. Curchod,
  N.~Ashari-Astani, I.~Tavernelli, U.~Rothlisberger, M.~K. Nazeeruddin,
  M.~Gr\"{a}tzel, Dye-sensitized solar cells with 13
  through the molecular engineering of porphyrin sensitizers, \emph{Nat Chem}
  (2014).

\bibitem{Bartok2017SA}
A.~P. Bart{\'o}k, S.~De, C.~Poelking, N.~Bernstein, J.~R. Kermode,
  G.~Cs{\'a}nyi, M.~Ceriotti, Machine learning unifies the modeling of
  materials and molecules, \emph{Sci. Adv.}, \textbf{3} (2017).

\bibitem{Sanchez-Lengeling2018S}
B.~Sanchez-Lengeling, A.~Aspuru-Guzik, Inverse molecular design using machine
  learning: Generative models for matter engineering, \emph{Science},
  \textbf{361}, 360--365 (2018).

\bibitem{Ahmad2016IJP}
I.~Ahmad, S.~Ahmed, Z.~Anwar, M.~A. Sheraz, M.~Sikorski, Photostability and
  Photostabilization of Drugs and Drug Products, \emph{Int. J. Photoenergy},
  \textbf{2016}, 1--19 (2016).

\bibitem{Koeppel84ACP}
H.~K{\"o}ppel, W.~Domcke, L.~S. Cederbaum, Multimode molecular dynamics beyond
  the Born-Oppenheimer approximation, \emph{Adv. Chem. Phys.}, \textbf{57},
  59--246 (1984).

\bibitem{Koeppel2001JCP}
H.~K\"{o}ppel, J.~Gronki, S.~Mahapatra, Construction scheme for regularized
  diabatic states, \emph{J. Chem. Phys.}, \textbf{115}, 2377--2388 (2001).

\bibitem{Worth2004ARPC}
G.~A. Worth, L.~S. Cederbaum, Beyond Born-Oppenheimer: Molecular Dynamics
  Through a Conical Intersection, \emph{Ann. Rev. Phys. Chem.}, \textbf{55},
  127--158 (2004).

\bibitem{Bowman2008MP}
J.~M. Bowman, T.~Carrington, H.~Meyer, Variational quantum approaches for
  computing vibrational energies of polyatomic molecules, \emph{Mol. Phys.},
  \textbf{106}, 2145--2182 (2008).

\bibitem{Meyer2009}
H.-D. Meyer, F.~Gatti, G.~A. Worth, \emph{Multidimensional Quantum Dynamics},
  Wiley-VCH Verlag GmbH \& Co. KGaA (2009).

\bibitem{Alborzpour2016JCP}
J.~P. Alborzpour, D.~P. Tew, S.~Habershon, Efficient and accurate evaluation of
  potential energy matrix elements for quantum dynamics using Gaussian process
  regression, \emph{J. Chem. Phys.}, \textbf{145}, 174112 (2016).

\bibitem{Richings2017CPL}
G.~W. Richings, S.~Habershon, Direct grid-based quantum dynamics on propagated
  diabatic potential energy surfaces, \emph{Chemical Physics Letters},
  \textbf{683}, 228 -- 233 (2017), ahmed Zewail (1946-2016) Commemoration Issue
  of Chemical Physics Letters.

\bibitem{Liu2017SR}
F.~Liu, L.~Du, D.~Zhang, J.~Gao, Direct Learning Hidden Excited State
  Interaction Patterns from ab initio Dynamics and Its Implication as
  Alternative Molecular Mechanism Models, \emph{Sci. Rep.}, \textbf{7}, 1--12
  (2017).

\bibitem{Richings2018JCP}
G.~W. Richings, S.~Habershon, {MCTDH} on-the-fly: Efficient grid-based quantum
  dynamics without pre-computed potential energy surfaces, \emph{J. Chem.
  Phys.}, \textbf{148}, 134116 (2018).

\bibitem{Williams2018JCP}
D.~M.~G. Williams, W.~Eisfeld, Neural network diabatization: A new ansatz for
  accurate high-dimensional coupled potential energy surfaces, \emph{J. Chem.
  Phys.}, \textbf{149}, 204106 (2018).

\bibitem{Xie2018JCP}
C.~Xie, X.~Zhu, D.~R. Yarkony, H.~Guo, Permutation invariant polynomial neural
  network approach to fitting potential energy surfaces. IV. Coupled diabatic
  potential energy matrices, \emph{J. Chem. Phys.}, \textbf{149}, 144107
  (2018).

\bibitem{Guan2019PCCP}
Y.~Guan, D.~H. Zhang, H.~Guo, D.~R. Yarkony, Representation of coupled
  adiabatic potential energy surfaces using neural network based quasi-diabatic
  Hamiltonians: {1,2 2A'} states of {LiFH}, \emph{Phys. Chem. Chem. Phys.},
  DOI:10.1039/C8CP06598E (2019).

\bibitem{Richings2019JCTC}
G.~W. Richings, C.~Robertson, S.~Habershon, Improved on-the-Fly MCTDH
  Simulations with Many-Body-Potential Tensor Decomposition and Projection
  Diabatization, \emph{J. Chem. Theory Comput.}, \textbf{15}, 857--870 (2019).

\bibitem{Polyak2019JCP}
I.~Polyak, G.~W. Richings, S.~Habershon, P.~J. Knowles, Direct quantum dynamics
  using variational Gaussian wavepackets and Gaussian process regression,
  \emph{J. Chem. Phys.}, \textbf{150}, 041101 (2019).

\bibitem{Guan2019JCP}
Y.~Guan, H.~Guo, D.~R. Yarkony, Neural network based quasi-diabatic
  Hamiltonians with symmetry adaptation and a correct description of conical
  intersections, \emph{J. Chem. Phys.}, \textbf{150}, 214101 (2019).

\bibitem{Wang2019JPCA}
Y.~Wang, C.~Xie, H.~Guo, D.~R. Yarkony, A Quasi-Diabatic Representation of the
  1,21A States of Methylamine, \emph{J. Phys. Chem. A}, \textbf{123},
  5231--5241 (2019).

\bibitem{Guan2020JCTC}
Y.~Guan, H.~Guo, D.~R. Yarkony, Extending the Representation of Multistate
  Coupled Potential Energy Surfaces To Include Properties Operators Using
  Neural Networks: Application to the 1,21A States of Ammonia, \emph{J. Chem.
  Theory Comput.}, \textbf{16}, 302--313 (2020).

\bibitem{Behler2008PRB}
J.~Behler, K.~Reuter, M.~Scheffler, Nonadiabatic effects in the dissociation of
  oxygen molecules at the {Al}(111) surface, \emph{Phys. Rev. B}, \textbf{77},
  115421 (2008).

\bibitem{Carbogno2010PRB}
C.~Carbogno, J.~Behler, K.~Reuter, A.~Gro\ss{}, Signatures of nonadiabatic
  ${\text{O}}_{2}$ dissociation at {Al}(111): First-principles fewest-switches
  study, \emph{Phys. Rev. B}, \textbf{81}, 035410 (2010).

\bibitem{Hu2018JPCL}
D.~Hu, Y.~Xie, X.~Li, L.~Li, Z.~Lan, Inclusion of Machine Learning Kernel Ridge
  Regression Potential Energy Surfaces in On-the-Fly Nonadiabatic Molecular
  Dynamics Simulation, \emph{J. Phys. Chem. Lett.}, \textbf{9}, 2725--2732
  (2018).

\bibitem{Dral2018JPCL}
P.~O. Dral, M.~Barbatti, W.~Thiel, Nonadiabatic Excited-State Dynamics with
  Machine Learning, \emph{J. Phys. Chem. Lett.}, \textbf{9}, 5660--5663 (2018).

\bibitem{Chen2018JPCL}
W.-K. Chen, X.-Y. Liu, W.-H. Fang, P.~O. Dral, G.~Cui, Deep Learning for
  Nonadiabatic Excited-State Dynamics, \emph{J. Phys. Chem. Lett.}, \textbf{9},
  6702--6708 (2018).

\bibitem{Westermayr2019CS}
J.~Westermayr, M.~Gastegger, M.~F. S.~J. Menger, S.~Mai, L.~Gonz\'alez,
  P.~Marquetand, Machine learning enables long time scale molecular
  photodynamics simulations, \emph{Chem. Sci.}, \textbf{10}, 8100--8107 (2019).

\bibitem{Westermayr2019arXiv}
J.~Westermayr, F.~A. Faber, A.~S. Christensen, O.~von Lilienfeld,
  P.~Marquetand, Neural networks and kernel ridge regression for excited states
  dynamics of {CH}$_2${NH}$_2^+$: From single-state to multi-state
  representations and multi-property machine learning models, arXiv:1912.08484
  [physics.chem--ph] (2019).

\bibitem{Ibele2019MP}
L.~M. Ibele, A.~Nicolson, B.~F.~E. Curchod, Excited-state dynamics of molecules
  with classically driven trajectories and Gaussians, \emph{Molecular Physics},
  DOI:10.1080/00268976.2019.1665199 (2019).

\bibitem{Tully1990JCP}
J.~C. Tully, Molecular Dynamics with Electronic Transitions, \emph{J. Chem.
  Phys.}, \textbf{93}, 1061--1071 (1990).

\bibitem{Richter2011JCTC}
M.~Richter, P.~Marquetand, J.~Gonz\'{a}lez-V\'{a}zquez, I.~Sola,
  L.~Gonz\'{a}lez, {SHARC:} {Ab} Initio Molecular Dynamics with Surface Hopping
  in the Adiabatic Representation Including Arbitrary Couplings, \emph{J. Chem.
  Theory Comput.}, \textbf{7}, 1253--1258 (2011).

\bibitem{Mai2018WCMS}
S.~Mai, P.~Marquetand, L.~Gonz\'alez, Nonadiabatic Dynamics: The SHARC
  Approach, \emph{WIREs Comput. Mol. Sci.}, \textbf{8}, e1370 (2018).

\bibitem{sharc-md2}
S.~Mai, M.~Richter, M.~Ruckenbauer, M.~Oppel, P.~Marquetand, L.~Gonz\'alez,
  SHARC2.0: Surface Hopping Including Arbitrary Couplings -- Program Package
  for Non-Adiabatic Dynamics, sharc-md.org (2018).

\bibitem{Doltsinis2006NIC}
N.~L. Doltsinis, \emph{Molecular Dynamics Beyond the Born-Oppenheimer
  Approximation: Mixed Quantum-Classical Approaches}, volume~31 of \emph{NIC
  Series}, John von Neuman Institut for Computing (2006).

\bibitem{Ha2018JPCL}
J.-K. Ha, I.~S. Lee, S.~K. Min, Surface Hopping Dynamics beyond Nonadiabatic
  Couplings for Quantum Coherence, \emph{J. Phys. Chem. Lett.}, \textbf{9},
  1097--1104 (2018).

\bibitem{Mai2020ACIE}
S.~Mai, L.~Gonz\'alez, Molecular Photochemistry: Recent Developments in Theory,
  \emph{Angew. Chem. Int. Edit.}, \textbf{n/a}.

\bibitem{Baer2002PR}
M.~Baer, Introduction to the theory of electronic non-adiabatic coupling terms
  in molecular systems, \emph{Phys. Rep.}, \textbf{358}, 75--142 (2002).

\bibitem{Lischka2004JCP}
H.~Lischka, M.~Dallos, P.~G. Szalay, D.~R. Yarkony, R.~Shepard, Analytic
  Evaluation of Nonadiabatic Coupling Terms at the {MR-CI} Level. {I.}
  {Formalism}, \emph{J. Chem. Phys.}, \textbf{120}, 7322--7329 (2004).

\bibitem{Rubbmark1981PRA}
J.~R. Rubbmark, M.~M. Kash, M.~G. Littman, D.~Kleppner, Dynamical effects at
  avoided level crossings: A study of the Landau-Zener effect using Rydberg
  atoms, \emph{Phys. Rev. A}, \textbf{23}, 3107--3117 (1981).

\bibitem{Hagedorn1991CMP}
G.~A. Hagedorn, Proof of the Landau-Zener formula in an adiabatic limit with
  small eigenvalue gaps, \emph{Communications in Mathematical Physics},
  \textbf{136}, 433--449 (1991).

\bibitem{Wittig2005JPCB}
C.~Wittig, The Landau-Zener Formula, \emph{J. Phys. Chem. B}, \textbf{109},
  8428--8430 (2005).

\bibitem{Zhu1995PRL}
L.~Zhu, V.~Kleiman, X.~Li, S.~P. Lu, K.~Trentelman, R.~J. Gordon, Ultrafast
  Coherent Control and Destruction of Excitons in Quantum Wells, \emph{Phys.
  Rev. Lett.}, \textbf{75}, 2598--2601 (1995).

\bibitem{Zhu2001JCP}
C.~Zhu, H.~Kamisaka, H.~Nakamura, Significant improvement of the trajectory
  surface hopping method by the Zhu–Nakamura theory, \emph{J. Chem. Phys.},
  \textbf{115}, 11036--11039 (2001).

\bibitem{Zhu2002JCP}
C.~Zhu, H.~Kamisaka, H.~Nakamura, New implementation of the trajectory surface
  hopping method with use of the Zhu–Nakamura theory. II. Application to the
  charge transfer processes in the 3D DH2+ system, \emph{J. Chem. Phys.},
  \textbf{116}, 3234--3247 (2002).

\bibitem{Kondorskiy2004JCP}
A.~Kondorskiy, H.~Nakamura, Semiclassical theory of electronically nonadiabatic
  chemical dynamics: Incorporation of the Zhu–Nakamura theory into the frozen
  Gaussian propagation method, \emph{J. Chem. Phys.}, \textbf{120}, 8937--8954
  (2004).

\bibitem{Oloyede2006JCP}
P.~Oloyede, G.~Mil’nikov, H.~Nakamura, Generalized trajectory surface hopping
  method based on the Zhu-Nakamura theory, \emph{J. Chem. Phys.}, \textbf{124},
  144110 (2006).

\bibitem{Ishida2017IRPC}
T.~Ishida, S.~Nanbu, H.~Nakamura, Clarification of nonadiabatic chemical
  dynamics by the Zhu-Nakamura theory of nonadiabatic transition: from
  tri-atomic systems to reactions in solutions, \emph{Int. Rev. Phys. Chem.},
  \textbf{36}, 229--285 (2017).

\bibitem{Granucci2012JCP}
G.~Granucci, M.~Persico, G.~Spighi, Surface hopping trajectory simulations with
  spin-orbit and dynamical couplings, \emph{J. Chem. Phys.}, \textbf{137},
  22A501 (2012).

\bibitem{Bartok2010PRL}
A.~P. Bart\'ok, M.~C. Payne, R.~Kondor, G.~Cs\'anyi, Gaussian Approximation
  Potentials: The Accuracy of Quantum Mechanics, without the Electrons,
  \emph{Phys. Rev. Lett.}, \textbf{104}, 136403 (2010).

\bibitem{Li2015PRL}
Z.~Li, J.~R. Kermode, A.~De~Vita, Molecular Dynamics with On-the-Fly Machine
  Learning of Quantum-Mechanical Forces, \emph{Phys. Rev. Lett.}, \textbf{114},
  096405 (2015).

\bibitem{Gastegger2015JCTC}
M.~Gastegger, P.~Marquetand, High-Dimensional Neural Network Potentials for
  Organic Reactions and an Improved Training Algorithm, \emph{J. Chem. Theory
  Comput.}, \textbf{11}, 2187--2198 (2015).

\bibitem{Rupp2015IJQC}
M.~Rupp, Machine learning for quantum mechanics in a nutshell, \emph{Int. J.
  Quantum Chem.}, \textbf{115}, 1058--1073 (2015).

\bibitem{Behler2016JCP}
J.~Behler, Perspective: Machine learning potentials for atomistic simulations,
  \emph{J. Chem. Phys.}, \textbf{145}, 170901 (2016).

\bibitem{Gastegger2016JCP}
M.~Gastegger, C.~Kauffmann, J.~Behler, P.~Marquetand, Comparing the accuracy of
  high-dimensional neural network potentials and the systematic molecular
  fragmentation method: A benchmark study for all-trans alkanes, \emph{J. Chem.
  Phys.}, \textbf{144}, 194110 (2016).

\bibitem{Gastegger2017CS}
M.~Gastegger, J.~Behler, P.~Marquetand, Machine learning molecular dynamics for
  the simulation of infrared spectra, \emph{Chem. Sci.}, \textbf{8}, 6924--6935
  (2017).

\bibitem{Deringer2017PRB}
V.~L. Deringer, G.~Cs\'anyi, Machine learning based interatomic potential for
  amorphous carbon, \emph{Phys. Rev. B}, \textbf{95}, 094203 (2017).

\bibitem{Botu2017JPCC}
V.~Botu, R.~Batra, J.~Chapman, R.~Ramprasad, Machine Learning Force Fields:
  Construction, Validation, and Outlook, \emph{J. Phys. Chem. C}, \textbf{121},
  511--522 (2017).

\bibitem{Smith2017CS}
J.~S. Smith, O.~Isayev, A.~E. Roitberg, ANI-1: an extensible neural network
  potential with DFT accuracy at force field computational cost, \emph{Chem.
  Sci.}, \textbf{8}, 3192--3203 (2017).

\bibitem{Behler2017ACIE}
J.~Behler, First Principles Neural Network Potentials for Reactive Simulations
  of Large Molecular and Condensed Systems, \emph{Angew. Chem. Int. Edit.},
  \textbf{56}, 12828--12840 (2017).

\bibitem{Zong2018npjCM}
H.~Zong, G.~Pilania, X.~Ding, G.~J. Ackland, T.~Lookman, Developing an
  interatomic potential for martensitic phase transformations in zirconium by
  machine learning, \emph{npj Comput Mater}, \textbf{4} (2018).

\bibitem{Bartok2018PRX}
A.~P. Bart\'ok, J.~Kermode, N.~Bernstein, G.~Cs\'anyi, Machine Learning a
  General-Purpose Interatomic Potential for Silicon, \emph{Phys. Rev. X},
  \textbf{8}, 041048 (2018).

\bibitem{Chmiela2018NC}
S.~Chmiela, H.~E. Sauceda, K.-R. M\"{u}ller, A.~Tkatchenko, Towards exact
  molecular dynamics simulations with machine-learned force fields, \emph{Nat.
  Commun.}, \textbf{9}, 3887 (2018).

\bibitem{Imbalzano2018JCP}
G.~Imbalzano, A.~Anelli, D.~Giofr\'e, S.~Klees, J.~Behler, M.~Ceriotti,
  Automatic selection of atomic fingerprints and reference configurations for
  machine-learning potentials, \emph{J. Chem. Phys.}, \textbf{148}, 241730
  (2018).

\bibitem{Zhang2018NIPS}
L.~Zhang, J.~Han, H.~Wang, W.~A. Saidi, R.~Car, E.~Weinan, End-to-end Symmetry
  Preserving Inter-atomic Potential Energy Model for Finite and Extended
  Systems, in: \emph{Proceedings of the 32Nd International Conference on Neural
  Information Processing Systems}, NIPS'18, 4441--4451, Curran Associates Inc.,
  USA (2018).

\bibitem{Zhang2018PRL}
L.~Zhang, J.~Han, H.~Wang, R.~Car, W.~E, Deep Potential Molecular Dynamics: A
  Scalable Model with the Accuracy of Quantum Mechanics, \emph{Phys. Rev.
  Lett.}, \textbf{120}, 143001 (2018).

\bibitem{Chan2019JPCC}
H.~Chan, B.~Narayanan, M.~J. Cherukara, F.~G. Sen, K.~Sasikumar, S.~K. Gray,
  M.~K.~Y. Chan, S.~K. R.~S. Sankaranarayanan, Machine Learning Classical
  Interatomic Potentials for Molecular Dynamics from First-Principles Training
  Data, \emph{J. Phys. Chem. C}, \textbf{123}, 6941--6957 (2019).

\bibitem{Christensen2019JCP}
A.~S. Christensen, F.~A. Faber, O.~A. von Lilienfeld, Operators in quantum
  machine learning: Response properties in chemical space, \emph{J. Chem.
  Phys.}, \textbf{150}, 064105 (2019).

\bibitem{Wang2019JCTC}
H.~Wang, W.~Yang, Toward Building Protein Force Fields by Residue-Based
  Systematic Molecular Fragmentation and Neural Network, \emph{J. Chem. Theory
  Comput.}, \textbf{15}, 1409--1417 (2019).

\bibitem{Chmiela2019CPC}
S.~Chmiela, H.~E. Sauceda, I.~Poltavsky, K.-R. M{\"{u}}ller, A.~Tkatchenko,
  sGDML: Constructing accurate and data efficient molecular force fields using
  machine learning, \emph{Comput. Phys. Commun.}, \textbf{240}, 38 -- 45
  (2019).

\bibitem{Carleo2019arXiv}
G.~Carleo, I.~Cirac, K.~Cranmer, L.~Daudet, M.~Schuld, N.~Tishby,
  L.~Vogt-Maranto, L.~Zdeborov\'a, Machine learning and the physical sciences
  (2019).

\bibitem{Krems2019PCCP}
R.~V. Krems, Bayesian machine learning for quantum molecular dynamics,
  \emph{Phys. Chem. Chem. Phys.}, \textbf{21}, 13392--13410 (2019).

\bibitem{Deringer2019AM}
V.~L. Deringer, M.~A. Caro, G.~Cs\'anyi, Machine Learning Interatomic
  Potentials as Emerging Tools for Materials Science, \emph{Advanced
  Materials}, \textbf{31}, 1902765 (2019).

\bibitem{Schuett2019NC}
K.~T. Sch\"{u}tt, M.~Gastegger, A.~Tkatchenko, K.-R. M\"{u}ller, R.~J. Maurer,
  Unifying machine learning and quantum chemistry with a deep neural network
  for molecular wavefunctions, \emph{Nat. Commun.}, \textbf{10}, 5024 (2019).

\bibitem{Schuett2018JCP}
K.~T. Sch\"{u}tt, H.~E. Sauceda, P.-J. Kindermans, A.~Tkatchenko, K.-R.
  M\"{u}ller, SchNet -- A deep learning architecture for molecules and
  materials, \emph{J. Chem. Phys.}, \textbf{148}, 241722 (2018).

\bibitem{Schuett2019JCTC}
K.~T. Sch\"utt, P.~Kessel, M.~Gastegger, K.~A. Nicoli, A.~Tkatchenko, K.-R.
  M\"uller, SchNetPack: A Deep Learning Toolbox For Atomistic Systems, \emph{J.
  Chem. Theory Comput.}, \textbf{15}, 448--455 (2019).

\bibitem{Haese2016CS}
F.~H\"ase, S.~Valleau, E.~Pyzer-Knapp, A.~Aspuru-Guzik, Machine learning
  exciton dynamics, \emph{Chem. Sci.}, \textbf{7}, 5139--5147 (2016).

\bibitem{Akimov2018JPCL}
A.~V. Akimov, A Simple Phase Correction Makes a Big Difference in Nonadiabatic
  Molecular Dynamics, \emph{J. Phys. Chem. Lett.}, \textbf{9}, 6096--6102
  (2018).

\bibitem{Bellonzi2019JCP}
N.~Bellonzi, G.~R. Medders, E.~Epifanovsky, J.~E. Subotnik, Configuration
  interaction singles with spin-orbit coupling: Constructing spin-adiabatic
  states and their analytical nuclear gradients, \emph{J. Chem. Phys.},
  \textbf{150}, 014106 (2019).

\bibitem{Lischka2001PCCP}
H.~Lischka, R.~Shepard, R.~M. Pitzer, I.~Shavitt, M.~Dallos, T.~M\"uller, P.~G.
  Szalay, M.~Seth, G.~S. Kedziora, S.~Yabushita, Z.~Zhang, High-Level
  Multireference Methods in the Quantum-Chemistry Program System {COLUMBUS:}
  Analytic {MR-CISD} and {MR-AQCC} Gradients and {MR-AQCC-LRT} for Excited
  States, {GUGA} Spin-Orbit {CI} and Parallel {CI} Density, \emph{Phys. Chem.
  Chem. Phys.}, \textbf{3}, 664--673 (2001).

\bibitem{Zhang2020JPCC}
Y.~Zhang, R.~J. Maurer, B.~Jiang, Symmetry-Adapted High Dimensional Neural
  Network Representation of Electronic Friction Tensor of Adsorbates on Metals,
  \emph{J. Phys. Chem. C}, \textbf{124}, 186--195 (2020).

\bibitem{Thiel1999JCP}
A.~Thiel, H.~K\"oppel, Proposal and numerical test of a simple diabatization
  scheme, \emph{J. Chem. Phys.}, \textbf{110}, 9371--9383 (1999).

\bibitem{Koeppel2006MP}
H.~K\"oppel, B.~Schubert, The concept of regularized diabatic states for a
  general conical intersection, \emph{Mol. Phys.}, \textbf{104}, 1069--1079
  (2006).

\bibitem{Maeda2010JCTC}
S.~Maeda, K.~Ohno, K.~Morokuma, Updated Branching Plane for Finding Conical
  Intersections without Coupling Derivative Vectors, \emph{J. Chem. Theory
  Comput.}, \textbf{6}, 1538--1545 (2010).

\bibitem{Kammeraad2016JPCL}
J.~A. Kammeraad, P.~M. Zimmerman, Estimating the Derivative Coupling Vector
  Using Gradients, \emph{J. Phys. Chem. Lett.}, \textbf{7}, 5074--5079 (2016).

\bibitem{Gonon2017JCP}
B.~Gonon, A.~Perveaux, F.~Gatti, D.~Lauvergnat, B.~Lasorne, On the
  applicability of a wavefunction-free, energy-based procedure for generating
  first-order non-adiabatic couplings around conical intersections, \emph{J.
  Chem. Phys.}, \textbf{147}, 114114 (2017).

\bibitem{Frankcombe2014JCP}
T.~J. Frankcombe, Using Hessian update formulae to construct modified Shepard
  interpolated potential energy surfaces: Application to vibrating surface
  atoms, \emph{J. Chem. Phys.}, \textbf{140}, 114108 (2014).

\bibitem{Richings2015IRPC}
G.~Richings, I.~Polyak, K.~Spinlove, G.~Worth, I.~Burghardt, B.~Lasorne,
  Quantum dynamics simulations using Gaussian wavepackets: the {vMCG} method,
  \emph{Int. Rev. Phys. Chem.}, \textbf{34}, 269--308 (2015).

\bibitem{An2018CPL}
H.~An, K.~K. Baeck, Practical and reliable approximation of nonadiabatic
  coupling terms between triplet electronic states using only adiabatic
  potential energies, \emph{Chem. Phys. Lett.}, \textbf{696}, 100 -- 105
  (2018).

\bibitem{Baeck2017JCP}
K.~K. Baeck, H.~An, Practical approximation of the non-adiabatic coupling terms
  for same-symmetry interstate crossings by using adiabatic potential energies
  only, \emph{J. Chem. Phys.}, \textbf{146}, 064107 (2017).

\bibitem{Plasser2019PCCP}
F.~Plasser, S.~Gómez, M.~F. S.~J. Menger, S.~Mai, L.~Gonz\'alez, Highly
  efficient surface hopping dynamics using a linear vibronic coupling model,
  \emph{Phys. Chem. Chem. Phys.}, \textbf{21}, 57--69 (2019).

\bibitem{Mai2019JCTC}
S.~Mai, A.~J. Atkins, F.~Plasser, L.~Gonz\'alez, The Influence of the
  Electronic Structure Method on Intersystem Crossing Dynamics. The Case of
  Thioformaldehyde, \emph{J. Chem. Theory Comput.}, \textbf{15}, 3470--3480
  (2019).

\bibitem{MOLPRO}
H.-J. {Werner \emph{et al.}}, MOLPRO, version 2012.1, A Package of Ab Initio
  Programs (2012), see http://www.molpro.net.

\bibitem{Barbatti2006MP}
M.~Barbatti, A.~J.~A. Aquino, H.~Lischka, Ultrafast two-step process in the
  non-adiabatic relaxation of the {CH$_{2}$NH$_{2}$} molecule, \emph{Mol.
  Phys.}, \textbf{104}, 1053--1060 (2006).

\bibitem{Larsen2017IOPP}
A.~H. Larsen, J.~J. Mortensen, J.~Blomqvist, I.~E. Castelli, R.~Christensen,
  M.~Du{\l}ak, J.~Friis, M.~N. Groves, B.~Hammer, C.~Hargus, E.~D. Hermes,
  P.~C. Jennings, P.~B. Jensen, J.~Kermode, J.~R. Kitchin, E.~L. Kolsbjerg,
  J.~Kubal, K.~Kaasbjerg, S.~Lysgaard, J.~B. Maronsson, T.~Maxson, T.~Olsen,
  L.~Pastewka, A.~Peterson, C.~Rostgaard, J.~Schi{\o}tz, O.~Sch\"utt,
  M.~Strange, K.~S. Thygesen, T.~Vegge, L.~Vilhelmsen, M.~Walter, Z.~Zeng,
  K.~W. Jacobsen, The atomic simulation environment{\textemdash}a Python
  library for working with atoms, \emph{J. Phys.: Condens. Matter},
  \textbf{29}, 273002 (2017).

\bibitem{Wigner1932PR}
E.~Wigner, On The Quantum Correction For Thermodynamic Equilibrium, \emph{Phys.
  Rev.}, \textbf{40}, 749--750 (1932).

\bibitem{Behler2015IJQC}
J.~Behler, Constructing high-dimensional neural network potentials: A tutorial
  review, \emph{Int. J. Quantum Chem.}, \textbf{115}, 1032--1050 (2015).

\bibitem{Granucci2010JCP}
G.~Granucci, M.~Persico, A.~Zoccante, Including quantum decoherence in surface
  hopping, \emph{J. Chem. Phys.}, \textbf{133}, 134111 (2010).

\bibitem{Feynman1939PR}
R.~P. Feynman, Forces in Molecules, \emph{Phys. Rev.}, \textbf{56}, 340--343
  (1939).

\bibitem{Herzberg1963DFS}
G.~Herzberg, H.~C. Longuet-Higgins, Intersection of potential energy surfaces
  in polyatomic molecules, \emph{Discuss. Faraday Soc.}, \textbf{35}, 77--82
  (1963).

\bibitem{Ryabinkin2017ACR}
I.~G. Ryabinkin, L.~Joubert-Doriol, A.~F. Izmaylov, Geometric Phase Effects in
  Nonadiabatic Dynamics near Conical Intersections, \emph{Acc. Chem. Res.},
  \textbf{50}, 1785--1793 (2017).

\bibitem{Dreuw2015WCMS}
A.~Dreuw, M.~Wormit, The algebraic diagrammatic construction scheme for the
  polarization propagator for the calculation of excited states, \emph{WIREs
  Comput. Mol. Sci.}, \textbf{5}, 82--95 (2015).

\bibitem{Richings2015TF}
G.~Richings, I.~Polyak, K.~Spinlove, G.~Worth, I.~Burghardt, B.~Lasorne,
  Quantum dynamics simulations using Gaussian wavepackets: the vMCG method,
  \emph{Int. Rev. Phys. Chem.}, \textbf{34}, 269--308 (2015).

\end{thebibliography}

\end{document}